\documentclass{aa}
\usepackage{graphicx}
\usepackage{epsfig}
\usepackage{natbib}
\usepackage{array}

\begin{document}

\title
{Synthetic hydrogen spectra of prominence oscillations }

\author
{ P. Heinzel$^1$ \and  M. Zapi\'{o}r$^{1, 2}$ \and  R. Oliver$^2$ \and J. L.
Ballester$^2$}

\offprints{J. L. Ballester\\ \email{joseluis.ballester@uib.es}}

\institute { Astronomical Institute, Academy of Sciences, 25165 Ond\v{r}ejov, The Czech Republic
\and
Departament de F\'{\i}sica, Universitat de les Illes Balears,
E-07122 Palma de Mallorca, Spain 
\email{pheinzel@asu.cas.cz; maciej.zapior@uib.es; ramon.oliver@uib.es; joseluis.ballester@uib.es}}

\date{Received 23 July 2013/ Accepted 21 November 2013}

\abstract {Prominence oscillations have been mostly detected using Doppler velocity, although there are also claimed detections by means 
of periodic variations in half-width or line intensity. However, scarce observational evidence exists about simultaneous detection of 
oscillations in several spectral indicators.}
 {Our main aim here is to explore the relationship between spectral indicators, such as Doppler shift, line intensity, and line half-width, and 
 the linear perturbations excited in a simple prominence model.}  
{Our equilibrium background model consists of a bounded, homogeneous
slab, which is permeated by a transverse magnetic field, having 
prominence-like physical properties. Assuming linear perturbations, the dispersion relation for fast and slow modes has been derived, 
as well as the perturbations for the different physical quantities. These perturbations have been used as the input variables in 
a one-dimensional radiative transfer code, 
which calculates the full spectral profile of the hydrogen H$\alpha$ and H$\beta$ lines.} 
{We have found that different oscillatory modes produce spectral indicator variations in different magnitudes.
Detectable variations in the Doppler velocity were found for the fundamental slow mode only. Substantial variations in the H$\beta$ line intensity were found for specific modes. Other modes lead to lower and even undetectable parameter variations.} 
{To perform prominence seismology, analysis of the H$\alpha$ and H$\beta$ spectral line parameters could be a good tool to detect and identify oscillatory modes.}
\keywords{Sun: oscillations -- Sun: filaments, prominences}
\titlerunning{Synthetic hydrogen spectra of prominence oscillations}
\authorrunning{Heinzel et al.}
\maketitle\maketitle

\section{Introduction}\label{intro}

Quiescent solar prominences are usually thought to be cool
($T \approx 10^{4}$\ K) and dense 
($\rho \approx  10^{-12} - 10^{-10}$ \ kg/m$^{3}$) plasma clouds located in less
dense and hotter solar corona.  It is still a matter of debate how these dense and cool structures live for a long time within the solar corona,
although a widely extended idea is that their support and thermal isolation are of magnetic
origin. They form along the inversion polarity line, or between the weak remnants of active regions. 
Prominences are highly dynamic structures subject to small amplitude oscillations, which have been routinely observed from the ground
and from space, and the details about their
properties can be found in \citet{oliver02, banerjee07, oliver09,
mackay10}  and \citet{arregui11a}. See also \cite{2012Arreui} for a comprehensive review of solar prominence oscillations. These small amplitude oscillations are characterized 
by one or more of the following features: They are not related to
flare activity; they display small amplitude velocities and are of local nature, where
only some regions of the prominence display periodic motions.
The most common technique to study prominence oscillations is to place a spectrograph slit on a prominence and to record one or more spectral 
lines along a time interval. The analysis of the obtained data provides the time series of several indicators such as Doppler velocity, line width, and 
line intensity, whose time behavior can be analyzed. \citet{tsubaki88} and \citet{oliver99} pointed out that in most of the papers devoted to 
observations of prominence oscillations, the periodicity is usually found in only one of the above mentioned spectral indicators. For instance, 
\citet{landman77} observed 
periodic fluctuations in the line intensity and width with period around 
22 min but not in the Doppler shift. \citet{yi91b} detected 
periods of 5 and 12 min in the 
power spectra of the line-of-sight velocity and the line intensity. Also, 
\citet{suematsu90} found signs of a $\sim$~60~min periodic variation in the 
Doppler velocity, line intensity, and line width. However, the Doppler 
signal also displayed shorter period variations (with periods around 4 
and 14 min), which were not present in the other two data sets. This
observation points out a striking feature already found in  other investigations, namely 
that the temporal behavior of various indicators corresponding to the 
same time series of spectra is not the same because either they show 
different periods in their power spectra \citep{tsubaki87} or one 
indicator presents a clear periodicity, while the others do not 
\citep{wiehr84, tsubaki86, balthasar86, tsubaki88b, suetterlin97}. 
Finally, \citet{balthasar94} 
simultaneously observed the spectral lines \ion{He}{i} 3888 \AA, H$_8$ 3889
\AA\, and 
\ion{Ca}{ii} 8498 \AA. From this information, they analyzed the temporal 
variations in the thermal and non-thermal line broadenings, the total H$_8$ 
line intensity, the \ion{He}{i} 3888 \AA\ to H$_8$ emission ratio, and the Doppler 
shift of the three spectral lines, which correlated well and thus reduced 
to a single data set. The power spectra of all these parameters yield a 
large number of power maxima, but only two of them (with periods of 29 and 
78 min) are present in more than one indicator.

The interpretation of the observational results summarized above appears to be difficult. These oscillations have been commonly interpreted 
in terms of standing or propagating magnetohydrodynamic (MHD) waves, and using this interpretation, a number of theoretical models have 
been set up to try to understand the prominence oscillatory behavior. Theoretical models can provide the temporal behavior of 
the plasma velocity, temperature, density, and other physical parameters, while observations yield information on quantities such as the 
line intensity or the line width and shift (line asymmetry). Therefore, a clear identification of  the spectral parameters with density, pressure, 
temperature, etc. is required before any progress can be achieved. Then, the presence of a certain period in more than one signal could be used to 
infer the properties of the MHD mode involved. Another source of useful information could be the detection of a 
given period in one signal but not in the others, as discussed above. 

Taking into account the above considerations, our main aim here is to explore the relationship between spectral indicators, such as Doppler 
shift, line intensity, and line half-width and the linear perturbations induced in a simple prominence model. The layout of the paper is as
follows: In Sect. 2, the equilibrium model and some theoretical
considerations are presented; in Sect. 3, we describe the method of solving the radiative-transfer
problem in oscillatory prominences; Sect. 4 contains our numerical results and their analysis.
Finally, our conclusions are drawn in Sect. 5.

\begin{figure}
\centering
\includegraphics[width=8.5cm]{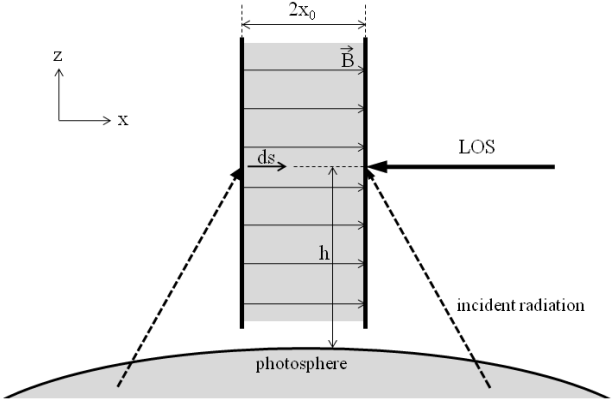}
\caption{Sketch of the geometry of the considered problem common for
  the MHD simulation and for the radiative transfer. The prominence slab with infinite heigth and length and width $2x_0$ is penetrated by a constant magnetic field. For a radiative transfer, we should take into account incident radiation from the photosphere (dashed arrows), according to the position above the solar limb ($h$). The line-of-sight is perpendicular to the slab.}   
\label{sketch}
\end{figure}

\begin{figure*}
\centering
\includegraphics[width=18.5cm, bb= 10 144 515 440]{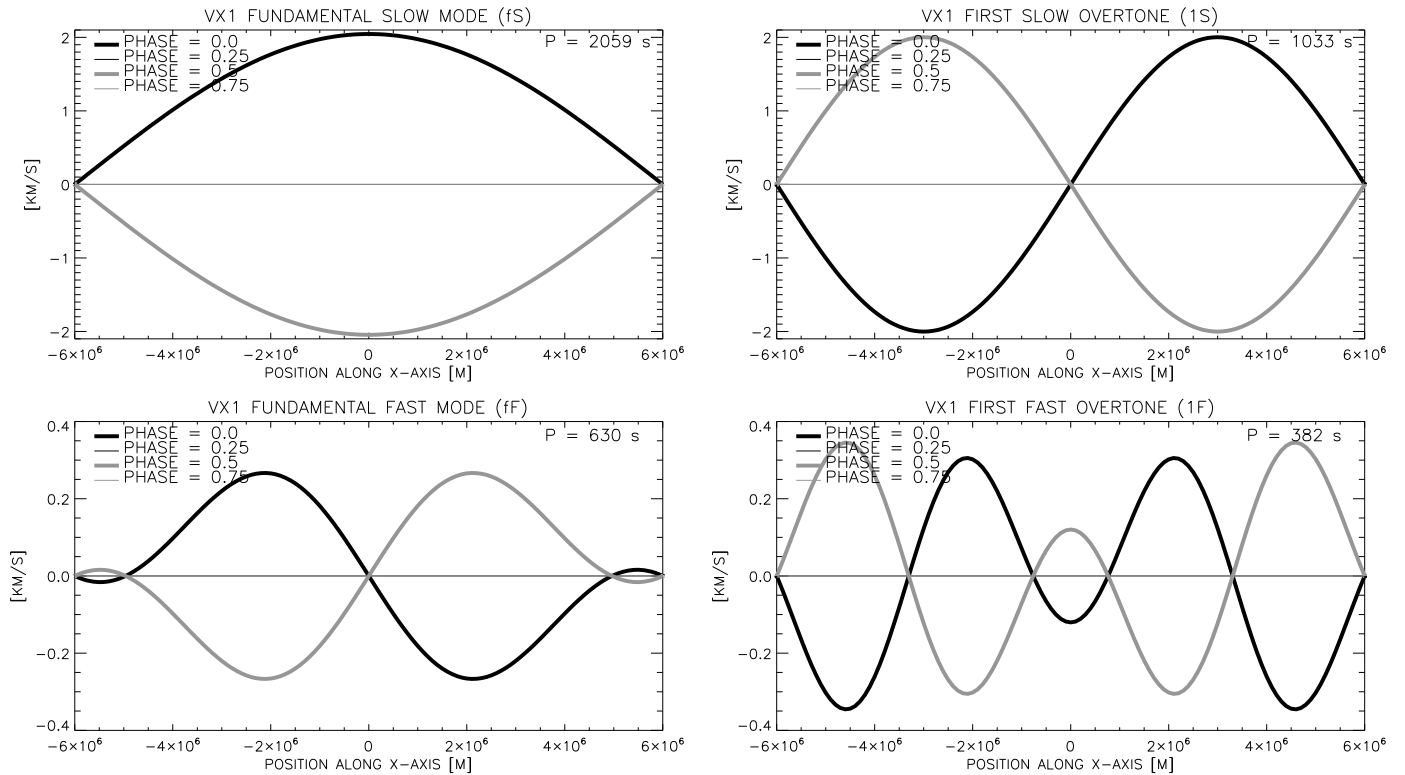}
\caption{Velocity perturbations along the $x$-axis of the slab for the fundamental slow mode (fS), the first slow overtone (1S), the fundamental fast mode (fF) and the first fast
overtone (1F). The $x$-axes of plots are scaled in meters.
Note the different velocity scale. Different lines represent different
phases (i.e. time divided by period), as labelled at the top left corner of each plot. 
For phase=0.25 and phase=0.75, the lines are merged. Oscillatory periods are shown at the top right corner of each plot.}   
\label{vx}
\end{figure*}

\begin{figure*}
\centering
\includegraphics[width=18.5cm]{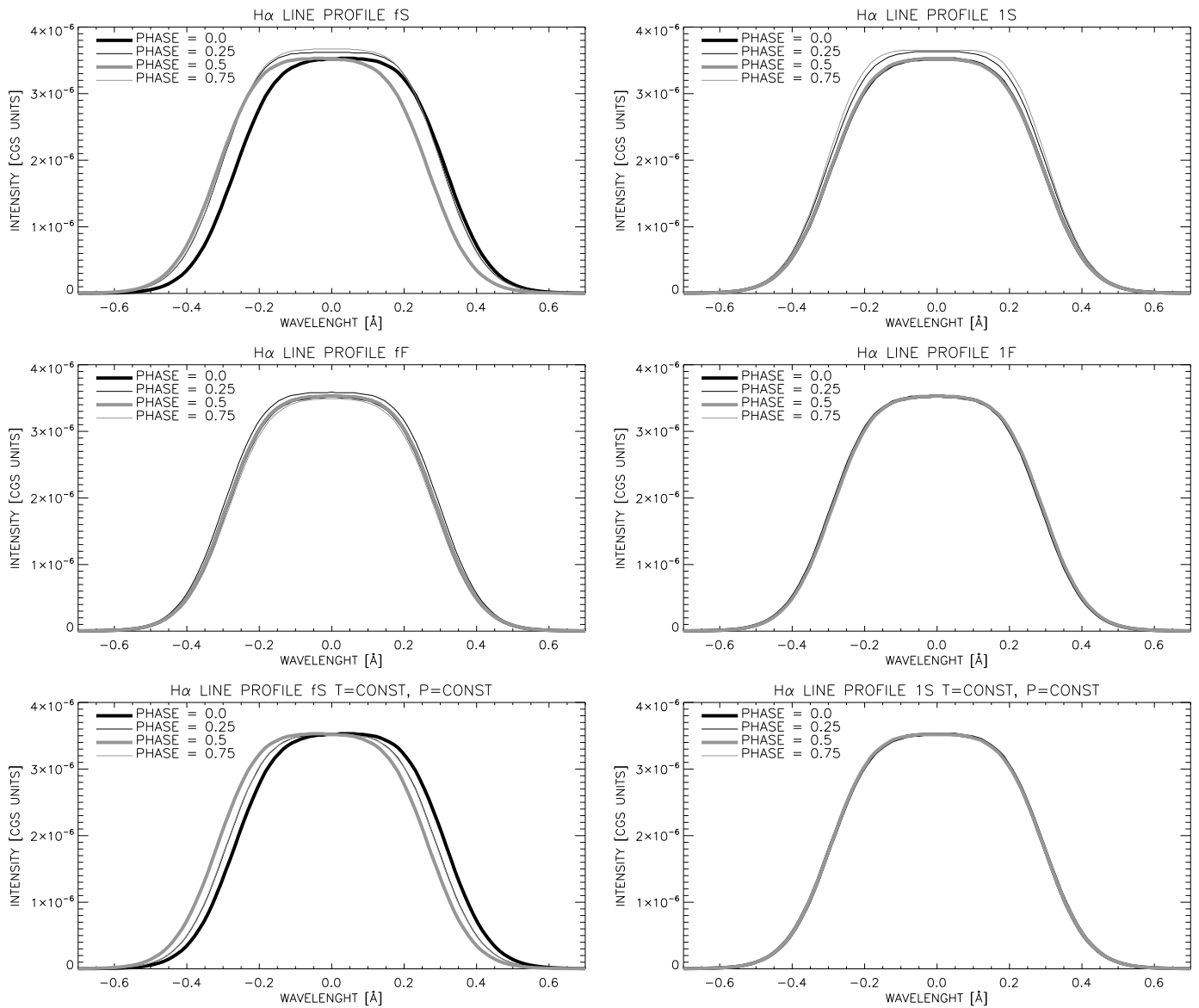}
\caption{Time variation in the H$\alpha$ spectral line profile for consecutive phases (labelled in the top left corner of each plot) and different modes.
The two bottom plots correspond to modes with $v_x$ variations only, where we assumed $T$=const. and $p$=const. Specific line intensities
are in cgs units erg sec$^{-1}$ cm$^{-2}$ sr$^{-1}$ Hz$^{-1}$.}   
\label{ha_prof}
\end{figure*}

\begin{figure*}
\centering
\includegraphics[width=18.5cm]{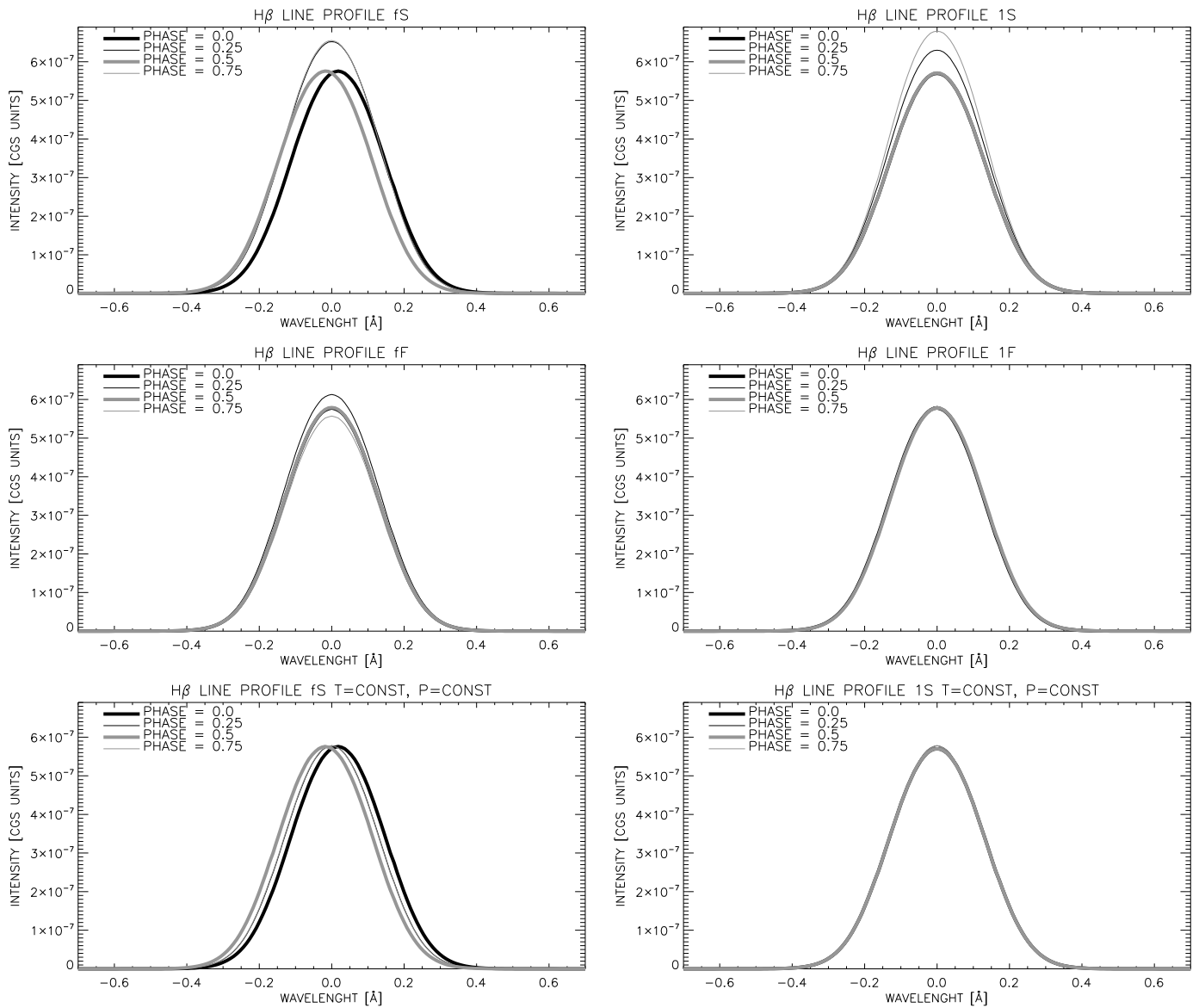}
\caption{Same as Fig. \ref{ha_prof}, but for H$\beta$.}   
\label{hb_prof}
\end{figure*}

\begin{figure*}
\centering
\includegraphics[width=18.5cm]{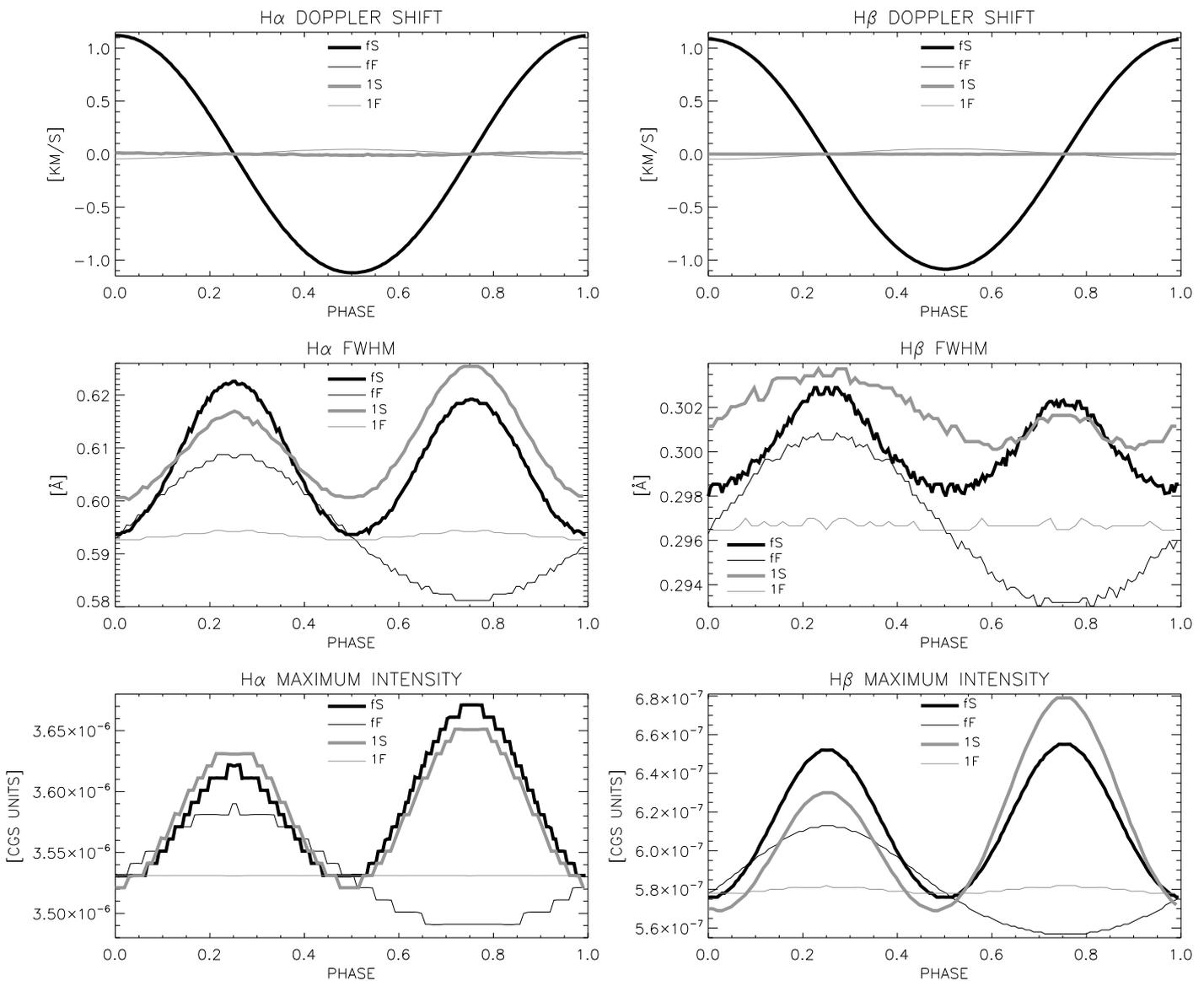}
\caption{Time variation as a function of phase of the spectral line parameters for all modes. Different lines labelled in the plots correspond 
to different modes. }   
\label{all}
\end{figure*}

\begin{figure*}
\centering
\includegraphics[width=18.5cm]{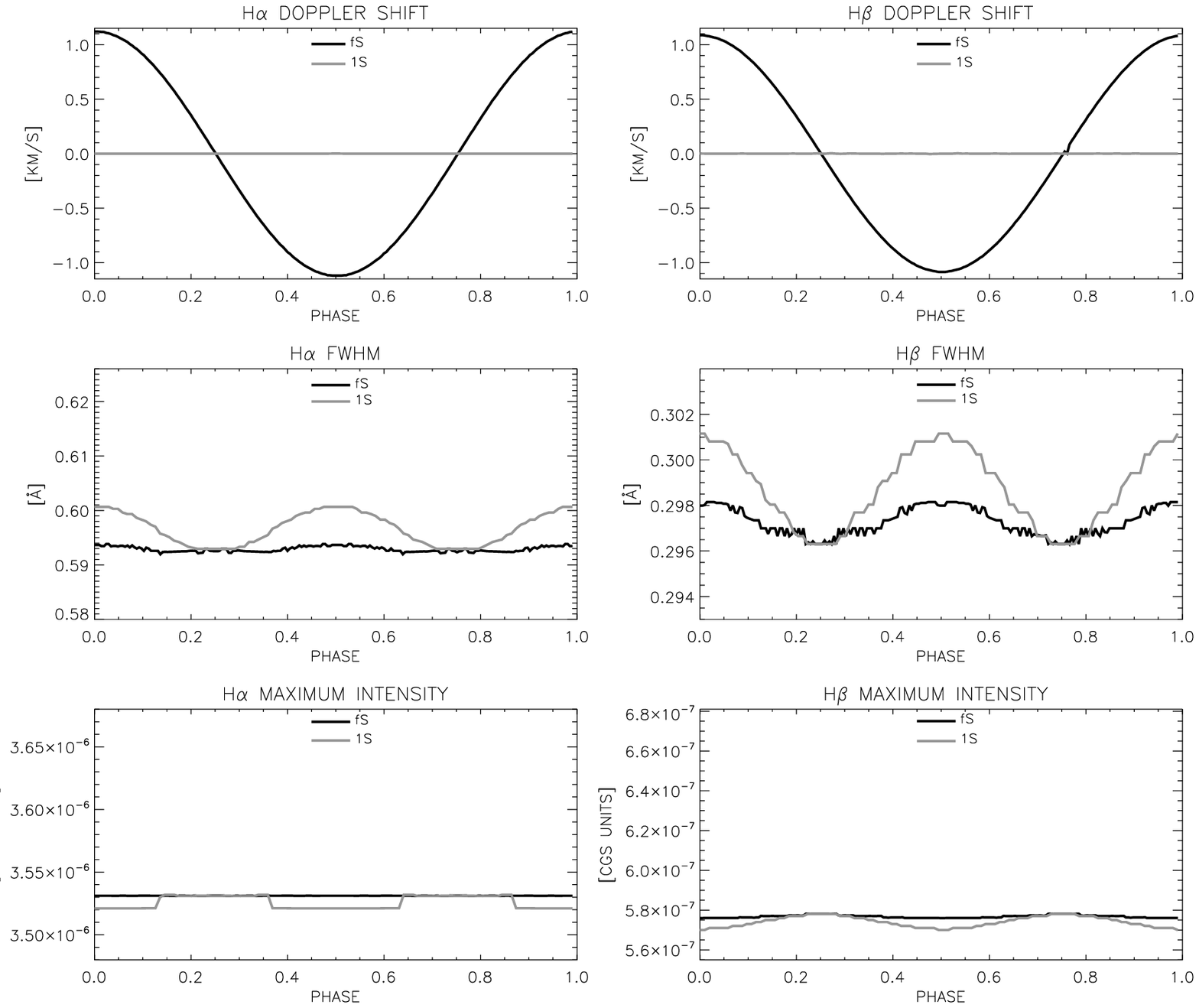}
\caption{Same as Fig. \ref{all}, but for $T$=const and $p$=const.}   
\label{all_test}
\end{figure*}

\section{Model and methods} \label{mod}

For our equilibrium configuration, we use a homogeneous slab bounded in
the transverse direction with a width $2x_{0}$ but of infinite length
in the direction along the slab, as well as infinite height (see
Figure \ref{sketch}). Such a slab is
threaded by a constant transverse magnetic field, which we assume to be
anchored in the lateral walls of the slab. The equilibrium magnitudes
of the slab are given by
\begin{eqnarray*}
     && p_0 = {\rm const.}, \ \ \rho_0 = {\rm const.}, \ \ T_0 = {\rm const.}, \\
     && {\bf
      B_0} =B_{0}\hat e_{x}, {\bf v}_0 = {\bf 0},
    \end{eqnarray*}
    with $B_{0}= 5\cdot10^{-4}$  \ T; $\rho_0 = 2.41 \cdot 10^{-10}$ kg m$^{-3}$; $\frac{T_0}{\tilde \mu} = 10^4$ K; 
    $p_0 = 0.02$ Pa, and $x_0$ =    6 $\cdot$ 10$^{6}$ \ m. Moreover, $\tilde \mu = 0.6$ for fully ionized plasma, which
    means that $T_0=6000$ K. 
With these parameters, the sound speed is
$$c_{\rm s} = \sqrt{\gamma R \frac{T_0}{\tilde \mu}}$$ \, ,
where $c_{\rm s}$ = 11761.5 m/s and the Alfv\'en speed is
 $$v_{\rm A} = \frac{B_{0}}{\sqrt{\mu_0 \rho_0}}$$
 ($v_{\rm A}$ = 63078.3 m/s). 

We consider linear and adiabatic MHD waves with small perturbations from the equilibrium in the form

    \begin{eqnarray*}
    && {\bf B}(t, {\bf r}) = {\bf B}_0 + {\bf B}_1 (t, {\bf r}), \\
    && p(t, {\bf r}) = p_0 + p_1(t, {\bf r}), \hspace{0.5cm} \rho(t, {\bf r}) = \rho_0 + \rho_1(t, {\bf r}) \\
    && T(t, {\bf r}) =
    T_{0} + T_{1}(t, {\bf r}), \ \ {\bf v}(t, {\bf r}) = {\bf v_1}(t, {\bf
    r}).
    \end{eqnarray*}

In addition, all perturbed variables, $f_1 (t, {\bf r})$, are taken as
\begin{equation}
f_1 (t, {\bf r})= f_1 (x) e^{i 
\left (\omega t  +
k_z z\right)}. \label{four}
\end{equation}

\noindent This corresponds to waves that propagate in the
$z$-direction along the prominence slab.
Considering only motions in the
$xz$-plane, the linearized MHD equations reduce to

\begin{equation} \label{13}
i \omega \rho_1 + \rho_0 {dv_x \over dx} + ik_z \rho_0 v_z = 0,
\end{equation}

\begin{equation} \label{14}
i \omega \rho_0 v_x = - {dp_1 \over dx},
\end{equation}

\begin{equation} \label{15}
i \omega \rho_0 v_z = - ik_z p_1 + {1 \over \mu} B_{0x} \left({dB_{1z}
\over dx} - i k_z B_{1x}\right),
\end{equation}

\begin{equation} \label{16}
i \omega p_1 - i \omega c^2_{\mathrm{s}} \rho_1 = 0,
\end{equation}

\begin{equation} \label{18}
i \omega B_{1x} = - ik_z B_{0x} v_z,
\end{equation}

\begin{equation} \label{19}
i \omega B_{1z} =B_{0x} {dv_z \over dx},
\end{equation}

\begin{equation} \label{20}
{p_1 \over p_0} = {\rho_1 \over \rho_0} + {T_1 \over T_0}.
\end{equation}
We proceed to eliminate all the 
perturbed quantities from the above equations, except for $v_{x}$ and $v_{z}$, and thus we end
up with the following set of coupled ordinary differential equations,

\begin{eqnarray}
c^2_{\mathrm{s}} {d^2 v_{x} \over dx^2} + \omega^2 v_x + ik_z  
c^2_{\mathrm{s}} {d 
v_{z} \over dx} = 0, \label{coupled1}
\end{eqnarray}
\begin{eqnarray}
 v^2_{\mathrm{A}} {d^2 v_{z} \over dx^2} + \left[\omega^2 - k^2_z 
 (v^2_{\mathrm{A}} + c^2_{\mathrm{s}})\right] v_{z} + ik_z c^2_{\mathrm{s}} {d v_{x} \over dx} = 0. \label{coupled2}
\end{eqnarray}
The above set of equations describes the coupled fast and slow magnetoacoustic modes in the considered equilibrium configuration, and 
Eqs.~(\ref{coupled1}) and (\ref{coupled2}) are the dimensional form of equations $(26)$ and $(27)$ in Sect. $3$ of \citet{oliver92}.
The procedure to obtain these solutions is presented in Sect. $3$ 
of \citet{oliver92}, while the method to obtain the dispersion
relations with  boundary conditions, which are given by
\begin{eqnarray*}
v_x(\pm x_0) = v_z(\pm x_0) = 0,
\end{eqnarray*}
is described in Sect. 4 of the same paper.

 Once $v_x(x)$ and $v_z(x)$ have been obtained, the rest of perturbations can be derived from the linearized equations ($\ref{13}) - (\ref{20}$). 
 Finally, these $x$-dependent expressions are combined with the exponential in Eq.~($\ref{four}$) and so we have the full spatial and temporal
 variations in all perturbed quantities.
 
On the other hand, to compute the synthetic spectra from model prominences with oscillations, one has to solve
the full NLTE radiative-transfer problem, which is
time-dependent in general (NLTE stands for departures from Local Thermodynamic
Equilibrium, LTE). The input model
is the MHD model of an oscillating prominence. Such a model can be pre-computed independently on
radiative-transfer modeling or computed consistently with it. The
latter case corresponds to a RMHD approach
(radiation MHD). While in the first case, one assumes or estimates the ionization degree of the plasma
(or simply of the hydrogen) and computes the radiation losses in an approximate way, in the RMHD approach
these quantities, which enter MHD equations, are consistent with the internal radiation field and excitation-ionization
state of the plasma. 

Here, we follow the first approach. From the pre-computed MHD model, we take the time variations in temperature $T$, gas
pressure $p$, and velocity along $x$-axis, $v_x$ (this is the
line-of-sight velocity in our case) as the main input for the
synthesis of spectral line profiles described in the next section. Since the perturbations 
 are written as $f(x,z,t)$, for our computations we have assumed $z = 0$, $x \in (-x_0, +x_0)$ and have computed the time-dependent perturbations 
 using time steps between 5 and 10 s. For times
$t=0, P/4, P/2, 3P/4$ and $P$ with $P$ being the oscillatory period, the behavior of the transverse velocity component is shown in Figure \ref{vx}. This figure contains
four panels that correspond to the fundamental slow mode (fS) and its first overtone (1S) together with the fundamental fast mode and its first
overtone (fF and 1F, respectively). 

On the other hand, it is also necessary to point out that the
pre-computed MHD model is based on the assumption of fully ionized
plasma, while the radiation transfer model solves the ionization
equilibrium. In principle, the consideration of partially ionized
plasma in the MHD model would produce a modification of the
perturbations used as an input for the NLTE modeling. However, it remains to be explored whether these differences are of substantial importance or not. This effect will be explored in a subsequent work.

\section{NLTE radiative transfer for the hydrogen plasma}


In the case of short oscillation periods when the period is comparable to radiative-relaxation times,
a fully time-dependent solution of the NLTE problem is needed. However, when the periods are large enough,
one can solve the NLTE radiative transfer for a series of stationary snapshots. In the following, we describe the latter approach in detail and postpone a fully time-dependent solution to future studies. 
In our model, we assume the same geometry as described in Section
\ref{mod} (see Figure \ref{sketch}) with $z$ axis perpendicular to the
solar surface. Since the period of prominence oscillations is
relatively long, we assume the statistical equilibrium 
at selected time snapshots and solve the time-independent NLTE transfer problem. Here we take a snapshot every
5 or 10 sec, depending on the actual period. The radiative-transfer equation is solved as the two-point boundary-value problem within our 1D slab.
Each side of the slab is irradiated by surrounding solar atmosphere, namely by photospheric and chromospheric
surface. This incident radiation is supposed to be the same on both slab surfaces. For static prominence models one
usually considers a symmetrical slab and solves the transfer problem only for half of it, specifying the boundary
condition in the slab center \citep[see e.g.][]{1995heinzel}. However, the oscillating slab is no longer symmetrical
because of the spatial distribution of various parameters. Therefore, we have to consider the full 1D slab
but with same boundary conditions on both its sides. Apart from this asymmetry of the slab, the solution to
NLTE problem is the same as described in \citet{1995heinzel}. We compute the excitation and ionization
balance for hydrogen consistently with the radiation field inside the slab, which is mainly determined by scattering
of the incident radiation. For medium gas pressures, as in this example, the thermal processes are less important, and
the partially-ionized plasma exhibits strong departures from LTE, typical of quiescent prominences. 
Another approximation we used
is the one for very small amplitudes of velocities (up to 2 km/sec in
present models), where one can solve the NLTE transfer
problem in two subsequent steps \citep{1998nejezchleba, 2005berlicki}. 
First, the static NLTE model is computed, and by using fixed atomic-level
populations and computed electron densities, we then perform the so-called 'formal solution' of the transfer equation,
where we include the line-of-sight velocity in computing the line opacities and emissivities (simply modifying
the respective profiles by accounting for the Doppler shifts). As a result, we obtain
the emergent emission profile of the studied spectral line for each time step (snapshot). This profile is in general asymmetrical due to velocities
or is Doppler shifted. 

Using the prescribed time-dependent MHD model, we performed this NLTE transfer modeling for a hydrogen model
atom having five bound levels and continuum. For a static case, this type of modeling is described in  \citet{1995heinzel}
and summarized in \citet{2010labrosse}. Present computations have been performed assuming the complete
redistribution in Lyman lines. We took the height of the line of sight above the limb to be 10000 km, which determines the dilution
of the incident radiation. For the sake of simplicity, we set the microturbulent velocity to zero; for more realistic values,
the line-center optical thickness ($\tau$) will decrease.

\section{Results}
We have obtained time-dependent series of generally asymmetrical line profiles of the hydrogen H$\alpha$ (Figure \ref{ha_prof}) and H$\beta$ lines (Figure \ref{hb_prof}).
H$\alpha$ is saturated (flat line core) because the line-center optical thickness is around 4, while H$\beta$ has an
almost Gaussian shape and is about an order of magnitude thinner, i.e. $\tau_0 < 1$ (the line-center
optical thickness).
For some other models with different temperatures
and gas pressures, a central reversal may be present in H$\alpha$ as
in the set of static models of \citet{1993gouttebroze}. 
From the set of profiles, we clearly see periodic oscillations
in both the line intensity (due to temperature and pressure variations) and in the wavelength position or
asymmetry. These variations have to be compared with spectral observations of oscillating prominences \citep[see][]{2012zapior}. 

We have to distinguish between two notions: the optical thickness and the optical depth. The optical 
depth is defined as
\begin{equation}
{\rm d}\tau(\lambda) = \kappa(\lambda) {\rm d}x ,
\end{equation}
where $\kappa(\lambda)$ is the wavelength-dependent absorption coefficient. In the case of quiescent
prominences, its profile is practically Gaussian but Doppler-shifted due to local oscillatory motions.
For the radiative-transfer modeling we transform the $x$-scale from $(-x_0, x_0)$ (shown in Fig. 1) to
$(0, 2x_0)$.
The optical depth is then
\begin{equation}
\tau(\lambda) = \int_{0}^{x} \kappa(\lambda) {\rm d}x'  .
\end{equation}
The optical thickness is simply the optical depth at $x=2x_0$, which corresponds to thickness
of the whole prominence slab. Both quantities vary from the line center towards the line
wings. In the line center, the slab is most opaque, while it becomes gradually
more and more transparent by going to the wings. This has principal consequences on the analysis of the synthetic (and also
observed) profiles because in the optically-thick regime (i.e. for $\tau(\lambda) > 1$) we can 'see' the
moving plasma only down to optical depths around unity. In cases when the slab is optically thin ($\tau(\lambda) < 1$),
we detect the motions from all depths, and the resulting line profile is a superposition of all depth-dependent
contributions. In the model studied here, the H$\alpha$ line-center optical thickness of the whole slab is around 4 in the
approximation of the complete frequency redistribution used here for simplicity and neglecting the 
microturbulent velocity, see discussion in Sec. 5.

The only evident asymmetry is detectable in H$\alpha$ for the fundamental slow mode (fS), and namely for phases 0. and 0.5 (see top left panel in Figure \ref{ha_prof}). We
clearly see the line asymmetry and a well resolved Doppler shift. 
For the first slow overtone (1S), these velocity-induced changes are much less evident and
the line profiles are dominated by thermodynamic perturbations.
To understand this behavior better, we
have made the following numerical experiment: we set all depth-dependent variations in temperature and gas pressure
to zero and considered a simple isothermal-isobaric slab characterised by the unperturbed thermodynamic quantities.
Then by using the same velocity fields as in the exact case, we calculated spectral profiles in the same way. The obtained profile variations with time for fS and 1S are plotted at the bottom
of Figure \ref{ha_prof}. While the fS shows similar Doppler effects as before (only the line center intensity is not varying), the 1S case
exhibits no variations. This clearly shows that the profile variations are significantly
affected by oscillations (always having the same sign at a given time)
in the case of fS. For 1S model the dominant role is
played by temperature/pressure changes and the effect of oscillatory motions is damped because of the antisymmetric
variations along $x$ (Figure \ref{vx}). The same behavior is seen for H$\beta$ in Figure \ref{hb_prof}. 
For fundamental fast mode (fF) and first fast overtone (1F), the
situation is even less favorable - we see only small profile variations due to temperature-pressure changes.    

From the computed line profiles of H$\alpha$ and H$\beta$, we derived three line parameters: Doppler shift, maximum intensity, and
FWHM (full width at half maximum). At the beginning, we approximate discrete points of the line profiles with the function
\begin{equation} 
I(\lambda, t) = S [1-{\rm e}^{-\tau_0 {\rm e}^{-{\left( \frac{\lambda - \lambda_0 + \Delta \lambda_V}{\Delta\lambda_D} \right)^2}}} ],
\end{equation}
where $I(\lambda)$ is the line profile, $S$ the source function, $\tau_0$ line-center optical thickness,  
$\lambda_0$ is the wavelength of the line center, $\Delta \lambda_V$  Doppler shift of the line profile, and $ \Delta \lambda_D$ is
the  Doppler width. This expression represents the formal solution of the transfer equation with a constant
line source function. We treated  $S$,  $\tau_0$ ,  $ \Delta \lambda_V$,  and  $ \Delta \lambda_D$ as free parameters. 
The maximum intensity of the spectral line was set to a peak value of the line profile. 
Since the spectral profiles were calculated with certain accuracy, the corresponding plots exhibit some noise. 
We also determined FWHM, where the  
noise is also caused by the numerical accuracy. 
Since the Doppler shift was calculated as a free parameter of the approximated curve, it is not so affected by such noise.
All parameter variations are shown in Figure \ref{all} and summarized in Tables \ref{tab:valuesa} and \ref{tab:valuesb}.
We detected substantial Doppler shifts for fS only with a peak-to-peak
value of about 2 km/s,  which is about a half the amplitude of plasma motions in the central part of the slab. 
The line FWHM shows slightly asymmetrical behavior in both lines and its amplitude is small. 
Finally, the
maximum line intensity is better detectable (variations about 10 \% and more) in the H$\beta$ line simply because H$\alpha$ is
saturated. 
We have done numerical test for fS and 1S modes with $T(t)=T_0$=const and $p(t)=p_0$=const. Results are plotted in Figure \ref{all_test}.
Doppler shifts are almost the same which means that they are not affected by variations in the thermodynamic parameters. However, 
FWHM and the maximum intensity look quite differently. This suggests that H$\beta$ maximum intensity can be
a good indicator of $T$ and $p$ variations.

Fast modes (fF and 1F) have a negligible Doppler shift, as expected
from a very low amplitude of $v_x$, and because the prominence axis is
perpendicular to the line-of-sight. Fast modes cause plasma motions
that are mainly in vertical direction parallel to the axis of the slab.
As mentioned before, the model 1S has also a negligible Doppler shift, which could be explained in the following way. 
The H$\alpha$ line center is optically thick, but a certain fraction of radiation comes from central parts of the slab in line wings. Since both the surface and
the slab center have zero velocities for 1S models, the contributions
to the Doppler signal from those regions are negligible. 
The models fS, f,F and 1S give rise of H$\alpha$ peak-to-peak FWHM variations in about 4-5 \% and 1-2\% for the H$\beta$. The peak-to-peak maximum intensity changes are about 3-4\%  for H$\alpha$ and 10--18\% for H$\beta$ also for modes fS, fF, and 1S. Especially for H$\beta$, these changes are detectable. Other cases are negligible. 

\renewcommand{\arraystretch}{1.25}

\begin{table}[t]
\caption{variations in spectral indicators for H$\alpha$.}
\centering
\begin{tabular}{cccc}
\hline
Mode & $\Delta v_{D}$ & FWHM & $\max[I(\lambda)]\times 10^{6}$ \\
\hline
fS     & 2.241 & 0.593 -- 0.622 (4.8\%) & 3.531 -- 3.671 (3.9\%)\\
1S     & 0.026 & 0.600 -- 0.625 (4.1\%) & 3.521 -- 3.651 (3.6\%)\\
fF     & 0.004 & 0.581 -- 0.609 (4.6\%) & 3.491 -- 3.590 (2.8\%)\\
1F     & 0.090 & 0.593 -- 0.594 (0.3\%) & 3.531 -- 3.531 (0.0\%)\\
fS - T & 2.240 & 0.592 -- 0.594 (0.3\%) & 3.531 -- 3.531 (0.0\%)\\
1S - T & 0.025 & 0.593 -- 0.601 (1.3\%) & 3.521 -- 3.532 (0.3\%)\\
\hline
\end{tabular}
\tablefoot{Abbreviations of modes are the same as described in the main text. T stands for numerical test with $T=$const and $p=$const. $\Delta v_{D}$ stands for peak-to-peak Doppler velocity amplitude (in km/sec) and FWHM is in \AA. In brackets, relative changes of each value with respect to the mean are presented.}
\label{tab:valuesa}
\end{table}

\begin{table}[t]
\caption{variations in spectral indicators for H$\beta$. }
\centering
\begin{tabular}{cccc}
\hline
Mode & $\Delta v_{D}$ & FWHM & $\max[I(\lambda)]\times 10^{7}$ \\
\hline
fS     & 2.175 & 0.298 -- 0.303 (1.6\%) & 5.760 -- 6.551 (12.8\%)\\
1S     & 0.005 & 0.300 -- 0.304 (1.2\%) & 5.690 -- 6.790 (17.6\%)\\
fF     & 0.018 & 0.293 -- 0.301 (2.6\%) & 5.570 -- 6.130 (9.6\%)\\
1F     & 0.098 & 0.296 -- 0.297 (0.2\%) & 5.780 -- 5.820 (0.7\%)\\
fS - T & 2.174 & 0.296 -- 0.298 (0.6\%) & 5.760 -- 5.820 (0.4\%)\\
1S - T & 0.007 & 0.296 -- 0.301 (1.6\%) & 5.700 -- 5.780 (1.4\%)\\
\hline
\end{tabular}
\tablefoot{See caption of Table \ref{tab:valuesa}.}
\label{tab:valuesb}
\end{table}

\section{Summary and future prospects}

In this study we have performed, for the first time, numerical simulations of the NLTE radiative transfer
in oscillating prominence slabs to investigate changes of the spectral
profiles. As the input, we used the MHD models that describe various modes of
global oscillations within simple 1D slabs. These slabs, which represent prominences as seen on the solar
limb, are strongly illuminated by the radiation from the solar disk and this radiation is scattered by the
prominence plasma. We have considered the hydrogen plasma and computed the synthetic spectral line
profiles of two Balmer lines, H$\alpha$ and H$\beta$, which are the most frequently observed optical lines
in prominences. The model considered here has typical temperatures and gas pressures, and we have
obtained a reasonable Doppler signal for the fundamental slow mode mode. Detected peak-to-peak amplitudes of velocities are of the order of 2 km/sec, while the real
one of the MHD model is 4 km/sec. The difference results from a response of the slab atmosphere to specific oscillations. 
Note that such velocities have been recently detected by \citet{2012zapior}.
We detected variations in the maximum intensity and FWHM for the fundamental slow mode and its first overtone and the fundamental fast mode, which are caused by temporal changes of the thermodynamic quantities, namely the temperature and gas pressure. 
Only the first fast overtone is practically non-detectable for all observables.

In this exploratory study, we did not consider the height variations (along the $z$-coordinate). 
To obtain a realistic distribution of the radiation field, atomic-level populations and electron densities,
one should solve the 2D transfer problem, as described for prominences by \citet{1995paletou} and \citet{2001heinzel_anzer}. 

The next step of the investigations may deal with the sensitivity of
the line profile to a particular velocity perturbation at a certain depth, which
is usually described by the so-called {\em velocity response function} \citep[e.g.][]{2000mein}.
In the future, a more rigorous treatment with the partial frequency redistribution (PRD) in hydrogen Lyman lines will be used. This leads to an increase in $\tau$ roughly by factor of two in the H$\alpha$ line center, as our test on current oscillatory models have indicated. However, non-zero microturbulent velocities
(not considered here) act in the opposite direction, lowering the optical thickness. Therefore, complete redistribution together with zero microturbulence
mimics more realistic situations somehow.
The presence of velocity field also makes the radiative transfer problem with the PRD more complex, and we thus postpone this issue to further studies.
The MHD model will be more realistic by considering immersion of the slab in a coronal environment with partially ionized plasma inside the slab. Partial ionization will also make it more consistent with the radiation transfer model we used in this study. 

The prominence magnetic field is
consistently incorporated into the MHD model but has no explicit effect on the synthetic line profiles.
However, an exploration of wider range of models and comparison to relevant observations could lead to
a realistic diagnostics of the magnetic field structure. Simultaneous polarimetric observations would then
be extremely important to compare the results, since the prominence seismology might serve as an 
independent diagnostics of the magnetic fields.
We have used only one specific initial temperature and gas pressure;
other thermodynamic models would lead to different results. 
For example, we could better disentangle between the surface and
internal dynamics of the slab by increasing the optical thickness
in H$\alpha$.
On the other hand, our model example is already opaque with $p$=0.02
Pa and the total
geometrical thickness $D$=12000 km (which is large compared to typical values for quiescent
filaments). Therefore, it is evident that some other spectral lines of other species, which are even
optically thicker than H$\alpha$ might be quite useful to diagnose the
oscillations. Namely, we have in mind
the \ion{Ca}{ii} H and K lines, which are optically thick and are easily detectable from the ground.
New perspectives also appear in connection with the IRIS space mission,
where the UV spectrometer covers strong and optically thick \ion{Mg}{ii} h and k lines \citep{2013Heinzel}.
Since IRIS provides a high temporal resolution, the prominence oscillations should be well
detectable.

\vspace{1cm}

{\bf Acknowledgements}
PH and MZ acknowledge the support by grant 209/12/0906 of the Grant Agency of the Czech Republic. This work was
also supported by the institute project RVO 67985815. 
MZ thanks the Astronomical Institute at Ond\v{r}ejov for the hospitality and support.
JLB, RO, and MZ acknowledge the financial support provided by MICINN and
FEDER funds under grant AYA2011-22846. JLB and RO acknowledge the financial support from CAIB and Feder Funds under the  \textquotedblleft Grups Competitius scheme.
\textquotedblright.
The authors thank the referee for valuable comments.

\bibliographystyle{aa}
\bibliography{Ref}

\end{document}